\documentstyle[multicol,aps,prb,epsf]{revtex}

\begin{document}



\title{
Enhancement of the d$_{x^2-y^2}$ pairing correlation in the 
two-dimensional Hubbard model: a quantum Monte Carlo study
}

\author{
Kazuhiko Kuroki and Hideo Aoki
}
\address{Department of Physics, University of Tokyo, Hongo,
Tokyo 113, Japan}

\date{\today}

\maketitle

\begin{abstract}
Quantum Monte Carlo is used to investigate the possibility of 
d$_{x^2-y^2}$ superconductivity in the two-dimensional 
repulsive Hubbard model.  A small energy scale 
relevant to possible pairing requires a care 
(i.e., sufficiently small level separation between the 
$k$ points 
$(\delta k,\pi-\delta k')$ and $(\pi-\delta k'',\delta k''')$ 
with small $\delta k$'s)
to detect enhanced correlations in finite-size studies, 
as motivated from a previous study on Hubbard ladders.  
Our calculation indeed detects 
long-tailed enhancements in the d$_{x^2-y^2}$ pairing correlation 
when the system is near, but not exactly at, half-filling.
\end{abstract}

\medskip


\begin{multicols}{2}
\narrowtext

\newpage
If the one-band Hubbard model, a simplest of the 
repulsively correlated electron systems, superconducts 
in two dimensions(2D), 
the interest is not only conceptually generic but may be practical 
as well, which has indeed been a challenge in the 
physics of high $T_C$ superconductivity.  
Some analytical calculations have suggested the occurrence
of d$_{x^2-y^2}$-wave superconductivity in the 2D Hubbard model. 
\cite{Bickers,Dzyalo,Schulz,Alvarez}
Numerical calculations have also been performed extensively.  
Finite binding energy\cite{DMST,FOP}
and pairing interaction vertex\cite{White,Husslein,Zhang} were
found in those calculations. 
Variational Monte Carlo calculations show that a superconducting 
order lowers the variational energy.\cite{Giamarchi,Yamaji2}
Quite recently, Hotta, Takada, and the present authors
have shown that the pairing with excluded double occupancies
has an enhanced correlation.\cite{KAHT}  
Nevertheless, there has been a reservation against the occurrence of 
superconductivity in the Hubbard model because 
the {\it bare} pairing correlation functions
do not show any symptom of long-range 
behavior.\cite{Zhang,Furukawa,Moreo}  

Now, quite a different avenue has emerged recently in the 
physics of ladders. While the weak-coupling theory for the two-leg 
Hubbard ladder \cite{Fabrizio,Schulz2,Balents} predicts 
the dominant pairing correlation,
the numerical calculations for small values of interaction 
gives enhancement of the pairing correlation only when 
the Fermi level for $U$(on-site repulsion)$=0$ 
lies between a bonding and an antibonding band 
levels that are separated with a sufficiently small level offset.
\cite{Yamaji,KTA,TKA}
This is because 
the relevant energy scale (the spin gap in the case of ladders) is so small 
that one has to make the separation between 
the highest occupied level (HOL) 
and the lowest unoccupied one (LUL) 
smaller than that to detect the pairing correlation in finite-size studies.  
The notion has also been successfully applied\cite{TKA}   
to confirm the weak-coupling prediction
that the three-leg Hubbard ladder can superconduct as well.
\cite{Schulz,TKA,Balents}

This view enables us to have a fresh look at the 2D Hubbard model.  
Interests here are two-fold: numerically, how will the 
QMC result behave when the care for a small LUL-HOL 
is taken.  Physically, will the pair-tunneling mechanism 
remain relevant also in 2D.
These are precisely the purpose of the present study.  
We have indeed found enhancements in 
the bare d$_{x^2-y^2}$ pairing correlation
in cases where the Fermi level lies between slightly separated LUL
$(\delta k, \pi-\delta k')$ and HOL $(\pi-\delta k'', \delta k)$,
between  which the pair tunneling should occur.  

It is instructive to start with the two-leg Hubbard ladder,
given as,
\begin{eqnarray*}
{\cal H}&=&-t_x\sum_{x,y,\sigma}
(c_{x,y,\sigma}^\dagger c_{x+1,y,\sigma}+{\rm h.c.})\nonumber\\&&
-t_y\sum_{x,\sigma}
(c_{x,1,\sigma}^\dagger c_{x,2,\sigma}+{\rm h.c.})
+U\sum_{x,y} n_{x,y,\uparrow}n_{x,y,\downarrow},
\end{eqnarray*}
where $x(=1,\cdots, N)$, $y(=1,2)$, and  $\sigma(=\uparrow,\downarrow)$ 
specifies the rungs, the chains, and the spins, respectively. 
According to the weak-coupling theory (perturbational renormalization +
bosonization), a gap opens in the spin excitations due to 
the relevance of interband pair tunneling 
between the Fermi points $(k_x, k_y) = (\pm k_F^0,0), (\pm k_F^{\pi},\pi)$. 
This concomitantly makes the two-point 
correlation of the interchain singlet, 
$c_{i,1,\uparrow}c_{i,2,\downarrow}-c_{i,1,\downarrow}c_{i,2,\uparrow}$, 
decay slowly with distance. 
In $k$-space the dominant component of this pair reads
\begin{equation}
\sum_{\sigma}\sigma\
(c_{k_F^0,\sigma}^0 c_{-k_F^0,-\sigma}^0-
c_{k_F^\pi,\sigma}^\pi c_{-k_F^\pi,-\sigma}^\pi),
\label{eqn1}
\end{equation}
where $c_{k,\sigma}^\mu$ annihilates an 
electron with spin $\sigma$ at $k_x=k$ in band $\mu(=0,\pi)$.  

Now, when the band structure is such that $E_F$ intersects 
the bonding-band top and the antibonding-band bottom with 
$k_F^0\sim\pi, k_F^\pi\sim 0$, 
{\it intrachain nearest-neighbor} singlet pair also has 
a dominant Fourier-component equal to eqn.(\ref{eqn1}) 
with a phase shift $\pi$ relative to the interchain
pairing. Thus, a linear combination, 
$
\sum_\sigma \sigma(c_{i,1,\sigma}c_{i,2,-\sigma}-
c_{i,y,\sigma}c_{i+1,y,-\sigma})
$
which amounts to the d$_{x^2-y^2}$ pairing, 
should have a slow decay as well.

Large enhancement of the inter-chain pairing correlation has in fact 
been found by exact diagonalization\cite{Yamaji} 
and by density matrix renormalization group\cite{Noack}
when $E_F$ lies close to the $k$-points 
$(0,\pi)$ and $(\pi,0)$.
Although the d$_{x^2-y^2}$-like nature of the pairing was 
suggested,\cite{Noack} d$_{x^2-y^2}$ pairing correlation 
itself has not been calculated.  
So, in our quest for 2D, 
we first calculate the correlation function with QMC.

Here we employ the ground-state, 
canonical-ensemble QMC,\cite{stab} where we take the free 
Fermi sea as the trial state. In most cases, we have 
taken the projection imaginary time $\tau$ to be $50/t_x$ or 
larger to ensure the 
convergence. We assume periodic boundary condition $c_{N+1}=c_1$.
We set $t_x=1$ hereafter.

We consider the case of 56 electrons in a $30\times 2$ lattice $(n=0.93)$
with $t_y=1.975$, where HOL is 
$(0,\pi)$ and LUL's are $(\pm 14\pi/15,0)$ 
for $U=0$. We have deliberately made 
the LUL deviate from exactly $(\pi,0)$ because,
although we have stressed that the interactions across 
$(\pi,0)$ and $(0,\pi)$ favors a d$_{x^2-y^2}$ pairing, 
the system becomes insulating 
if $E_F$ at $U=0$ (denoted by $E_F^0$ hereafter) lies exactly
between $(\pi,0)$ and $(0,\pi)$, 
for which the equality $k_F^0+k_F^{\pi}=\pi$ 
brings about the interband umklapp processes.\cite{Balents,KTA}

\begin{figure}
\begin{center}
\leavevmode\epsfysize=70mm \epsfbox{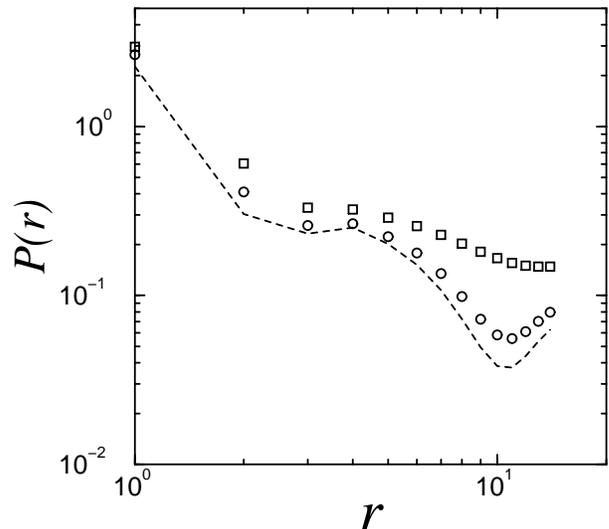}
\caption{
QMC result for the d$_{x^2-y^2}$ pairing correlation, $P(r)$, 
in a $30\times 2$ Hubbard ladder with 56 electrons with 
$U=1$ and $t_y=1.975$ $(\Box)$ or $t_y=1.93$ $(\bigcirc)$. 
The error bars are smaller than the symbols.
The dashed line represents the noninteracting case.
}
\label{fig1}
\end{center}
\end{figure}

In Fig.\ref{fig1}, we show the d$_{x^2-y^2}$ 
pairing correlation $P(r)$ defined by
$P(r)=\sum_{|\Delta x|+|\Delta y|=r} 
\langle O^\dagger (x+\Delta x,y+\Delta y) O (x,y)\rangle,$
where
$
O(x,y)=\sum_{\delta=\pm 1,\sigma}
\sigma(c_{x,y,\sigma} c_{x+\delta,y,-\sigma}
-c_{x,y,\sigma} c_{x,y+\delta,-\sigma}),
$
with $c_{x,3}\equiv c_{x,1}, c_{x,0}\equiv c_{x,2}$.
$P(r)$ for $U=1$ has an overall enhancement over the 
noninteracting result, and decays slowly.

Here we have tuned $t_y$ to make the LUL-HOL gap 
(denoted by $\Delta\varepsilon^0$ hereafter) 
as small as $0.006$. 
This procedure is important, at least for small $U$,
in detecting an enhanced pairing correlation 
in finite systems, 
as stressed in ref.\onlinecite{Yamaji} and in 
our previous publications.\cite{KTA,TKA} 
We can in fact see this 
in Fig.\ref{fig1}, where a 2 \% change into $t_y=1.93$ 
washes out the enhancement. This may seem surprisingly sensitive, but 
the change in $t_y$ is accompanied by a more than one order of 
magnitude increase in 
$\Delta\varepsilon^0 \rightarrow 0.1$ 
(without changing the $U=0$ ground state).  

Now we are in position to move on to the 2D Hubbard model.  
Here we consider the isotropic case of $t_y\simeq t_x$
and $x,y=1,\cdots,N$
with periodic boundary condition in both directions,
where the $k$-points around $(0,\pi)$ and $(\pi,0)$ are close in
energy. Our expectation from the study on ladders is that the 
pair tunneling processes between 
$(\delta k,\pi-\delta k')$ and $(\pi-\delta k'',\delta k''')$ 
may result in d$_{x^2-y^2}$ pairing, $\sum_{\bf k} [\cos(k_x)-\cos(k_y)] 
c_{{\bf k}\uparrow}c_{{-\bf k}\downarrow}$ in 2D,
but an enhanced pairing correlation
may be detected only when $\Delta\varepsilon^0$ between
those levels is small.
Apart from such an argument,
the importance of the interactions around 
$(0,\pi)$ and $(\pi,0)$
in the 2D Hubbard model has been suggested by 
various authors.
\cite{Dzyalo,Schulz,Alvarez,Husslein,Yamaji2,LeeRead,Newns,Mark,Gonzalez}

In the present approach, 
one should be able to detect the effect of a spin (or superconducting)
gap due to the tunneling processes 
in the situation where $E_F^0$ situates between 
closely lying levels of $(\delta k,\pi-\delta k')$ and 
$(\pi-\delta k'',\delta k''')$, where $\delta k$'s are small.

We first take
46 electrons in $8\times 8$ sites ($n=0.72$)
with $t_y=0.999$.  
For this band filling, $E_F^0$ lies between the levels of
$(0,\pm 3\pi/4)$ and $(\pm 3\pi/4,0)$.
We have taken $t_y=0.999$, because the number of electrons considered 
here would have an open shell (with a degeneracy in the free-electron 
Fermi sea) for $t_y=1$, which will destabilize QMC convergence. 
Taking $t_y=0.999$ lifts 
the degeneracy between $(k_1,k_2)$ and $(k_2,k_1)$ 
to give a tiny ($<0.01$) but finite $\Delta\varepsilon^0$.

\begin{figure}
\begin{center}
\leavevmode\epsfysize=70mm \epsfbox{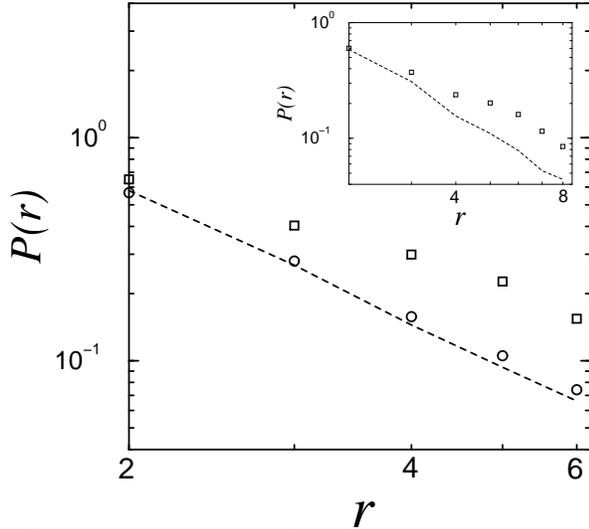}
\caption{
A plot similar to Fig.\protect\ref{fig1} for 
an $8\times 8$ system with 46 electrons, and 
$10\times 10$ sites with 78 electrons (inset).
$U=1$ and $t_y=0.999$ $(\Box)$ or $t_y=0.95$ ($\bigcirc$).
}
\label{fig2}
\end{center}
\end{figure}

In Fig.\ref{fig2} we plot the d$_{x^2-y^2}$ pairing correlation, where 
the correlation for $U=1$ is clearly seen to be enhanced over that for $U=0$ 
especially at large distances, resulting in a slower decay. 
A similar result is shown in the inset for a larger system 
($10\times 10$ with 78 electrons), where $E_F^0$ lies between 
$(0,\pm 4\pi/5)$ and $(\pm 4\pi/5,0)$.  
Also shown in Fig.\ref{fig2}(a) is 
a result for $t_y=0.95$, where $\Delta\varepsilon^0$ blows up to $\sim 0.17$, 
and the enhancement vanishes in accordance with the 
expectations from the ladders. 

\begin{figure}
\begin{center}
\leavevmode\epsfysize=70mm \epsfbox{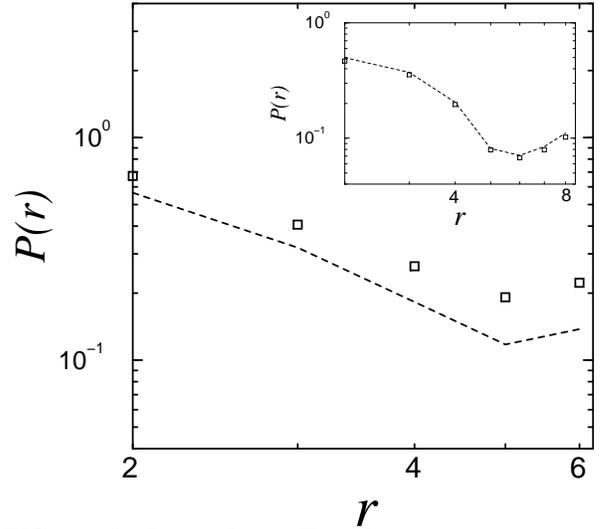}
\caption{
A plot similar to Fig.\protect\ref{fig2} for 
an $8\times 8$ system with 60 electrons, $t_y=0.999$, $t'=0$,
and a $10\times 10$ system with 74 electrons, $t_y=1$, $t'=-0.24$ (inset).
In both cases $U=0$ (dashed line) or $U=1 (\Box)$.
}
\label{fig3}
\end{center}
\end{figure}

In the above situation, 
LUL and HOL are taken to be 
$(0,\pi-\delta k)$ and $(\pi-\delta k,0)$ with 
$\Delta\varepsilon^0 < 0.01$, while the other 
levels lie more than $\sim 0.1$ away from $E_F^0$.  
One might thus raise a criticism that the scattering processes involving 
the states away from $E_F^0$ are unduly neglected.  
Some of these processes may favor the pairing, while 
others may not.
We can in fact focus on the scattering processes that
do not favor d$_{x^2-y^2}$ pairing by taking a LUL
state on the $\Gamma$-M ($|k_x|=|k_y|$) line, and situate all the others 
except a certain HOL away from $E_F^0$. 
There, 
d$_{x^2-y^2}$ pairing should not to be favored since the 
d$_{x^2-y^2}$ gap function has nodal lines along $\Gamma$-M. 
To realize such a situation, we introduce the 
next nearest-neighbor hopping
$t'=-0.24$\cite{tcomment} in a $10\times 10$ lattice with 74 
electrons where we take $t_y=1$ this time.\cite{Hcomment}
$E_F^0$ now lies between LUL $(\pm 2\pi/5, \pm 2\pi/5)$ 
along $\Gamma$-M line and HOL $(\pm4\pi/5,0)$, with
$\Delta\varepsilon^0\sim 0.01$.
\noindent
We can see from the inset of Fig.\ref{fig3} that the d$_{x^2-y^2}$ 
correlation is indeed no longer enhanced for $U=1$.

\begin{figure}
\begin{center}
\leavevmode\epsfysize=70mm \epsfbox{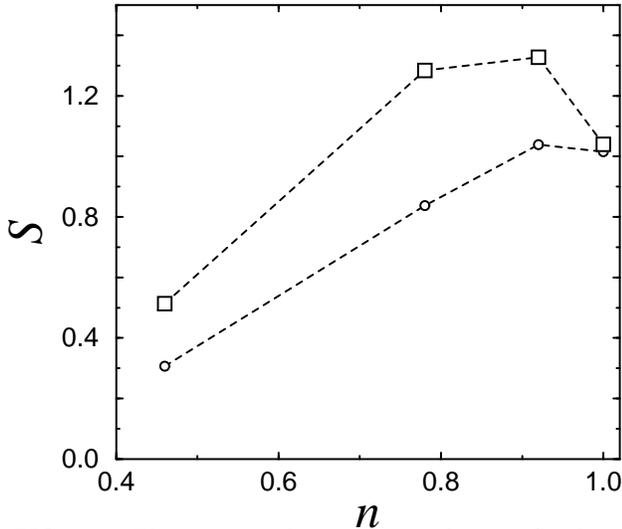}
\caption{
The integrated pairing correlation, $S$, 
plotted against band filling $n$ for
a $10\times 10$ lattice with $t_y=0.999$.
$U=0$ $(\bigcirc)$ or $U=1 (\Box)$.
}
\label{fig4}
\end{center}
\end{figure}

We have then to investigate the case where both the scatterings which 
do and do not favor d$_{x^2-y^2}$ pairing are taken into account
on an equal footing.
To realize such a situation, we can in fact take an $8\times 8$ lattice 
with 60 electrons with $t_y=0.999$ and $t'=0$. 
In this case, the energies of the 
occupied levels $(0,\pi)$ and $(\pm\pi/4,\pm 3\pi/4)$ are 
within 0.01 to those of the unoccupied levels $(\pi,0)$, 
$(\pm 3\pi/4,\pm \pi/4)$, and $(\pm\pi/2,\pm\pi/2)$,
so that the scatterings
between, e.g., $(0,\pi)$ and $(3\pi/4,\pi/4)$, which favors d$_{x^2-y^2}$
pairing,\cite{comment} and those between, e.g. ,
$(0,\pi)$ and $(\pi/2,\pi/2)$, which do not, coexist.
In Fig.\ref{fig3}(b) we can see that the d$_{x^2-y^2}$ correlation for $U=1$ 
is enhanced over the non-interacting result, 
so that the coexistence is not detrimental.  
A similar $k$-space configuration may be realized by taking 92 electrons 
in a $10\times 10$ system ($n=0.92$), for which an enhanced 
d$_{x^2-y^2}$ correlation is obtained as well (Fig.\ref{fig4}).

Our final important finding is the band-filling dependence. We have 
calculated 
$S \equiv \sum_{r\geq 3}P(r)$, which is a measure of the long-range 
part of the correlation, 
for the $10\times 10$ lattice with $t_y=0.999$ and 
$t'=0$ for various $n=1,\:0.92,\:0.78$, and $0.46$. 
$\Delta\varepsilon^0$ is kept to be $<0.01$ throughout. 
The summation is restricted to 
$r\geq 3$ in order to eliminate short-range contributions.
For $n=0.46$ (46 electrons) LUL/HOL is 
($(0,\pm 3\pi/5)$ and $(\pm 3\pi/5,0)$), which significantly deviate from 
$(0,\pi)$ and $(\pi,0)$.  
The result, displayed in Fig.\ref{fig4}, 
shows that the enhancement in $S$ for $U=1$ has a maximum around 
a finite doping. 
Although some enhancement over the $U=0$ result remains 
at $n=0.46$, the absolute value of $S$ becomes small reflecting the 
rounded-off Fermi surface for small $n$.  
Thus the message here is 
that the d$_{x^2-y^2}$ pairing is favored {\it near, but 
not exactly at}, half-filling.

To conclude, we have obtained an enhancement of d$_{x^2-y^2}$-wave
pairing correlation in the 2D Hubbard model.
Special care has been taken on $\Delta\varepsilon^0$ between the HOL
$(\delta k,\pi-\delta k')$ and LUL $(\pi-\delta k'',\delta k''')$, 
as motivated from the studies on ladders.
Analytical approaches for the 2D Hubbard model 
show that the relevant energy scale
for superconductivity, $\Delta_S$, is 
$O(0.01t)$.\cite{Bickers,Alvarez}  
In this light it is natural that $\Delta\varepsilon^0$ has to 
be smaller than $\sim O(0.01t)$ 
if one wishes to detect enhanced pairing correlations 
in finite systems.

Although the d$_{x^2-y^2}$ correlation is shown to be enhanced only when 
$\Delta\varepsilon^0$ is small, 
this result does not necessarily imply that a high density of states 
around the Fermi level is a prerequisite to 
superconductivity in the {\it thermodynamic limit}.  
Rather, we argue that the `high density of states' 
is necessitated in finite systems to make the situation 
closer to the thermodynamic 
limit, where $\Delta_S/\Delta \varepsilon^0$ diverges after all 
for any value of $D(E_F)$.
On the other hand, by preferentially focusing on the scatterings between,
e.g., $(0,\pi-\delta k)$ and $(\pi-\delta k,0)$ as in Fig.\ref{fig2}, 
we may be {\it mimicking} the 
situation in many of the high $T_C$ cuprates for which the photoemission
studies have revealed that $E_F$ lies 
very close to the extended van Hove singularity,\cite{Shen} because 
the scatterings involving the $k$-points around 
$(0,\pi)$ and $(\pi,0)$ would indeed be dominant 
due to the high density of states.

Present study is restricted to relatively small values of $U/t \sim 1$.
For large $U/t$, the strategy of paying attention
to the $U=0$ energy levels can fail to detect the enhancement
of the pairing correlation, if any, because 
free-electron levels may become irrelevant.  
Namely, some energy scale $\Delta \varepsilon ^{\rm eff}$
(possibly a kind of charge excitation energy),
which vanishes in the thermodynamic limit with
$\Delta_S/\Delta \varepsilon ^{\rm eff}\rightarrow\infty$, 
should still exist in finite systems even 
\noindent
for large $U/t$, but $\Delta\varepsilon ^{\rm eff} $
will not be dominated by $\Delta\varepsilon^0$.
Then the problem will be how to tune $\Delta \varepsilon ^{\rm eff}$.

Our results suggest that 
the pair-tunneling processes
are enhanced in the 2D Hubbard model,
which possibly leads to superconductivity.
Whether such an enhancement is 
due to, e.g., spin fluctuations
remains to be an open question.

Numerical calculations were performed at the Supercomputer Center,
Institute for Solid State Physics, University of Tokyo,
and at the Computer Center of the University of Tokyo.
This work was also supported in part by Grant-in-Aids for Scientific
Research from the Ministry of Education
of Japan.


\end{multicols}

\begin{references}
\bibitem{tcomment} We are not arguing that 
$t'$ always destroys d$_{x^2-y^2}$ pairing. 
In fact, d$_{x^2-y^2}$ pairing correlation is enhanced if
$E_F^0$ lies between $(\delta k,\pi-\delta k')$
and $(\pi-\delta k'',\delta k''')$, 
regardless of the presence/absence of $t'$, as will be published elsewhere.
\bibitem{Hcomment}This is close to the parameter values 
for which the vertex part of
the d$_{x^2-y^2}$ correlation was found to have a 
plateau-like behavior in ref.\protect\onlinecite{Husslein}. 
Since we do not find an enhancement of the 
{\it bare} d$_{x^2-y^2}$ pairing correlation for this parameter set, 
our results should not be confused with those 
in ref.\protect\onlinecite{Husslein},
where the behavior of the {\it vertex part} is 
used as the criterion for superconductivity.
\end{references}
\end{document}